# A POSSIBLE MECHANISM FOR THE ORIGIN OF INERTIA IN DE SITTER-FANTAPPIÉ-ARCIDIACONO PROJECTIVE RELATIVITY

**Leonardo Chiatti**

AUSL VT Medical Physics Laboratory
Via Enrico Fermi 15
01100 Viterbo (Italy)
fisica1.san@asl.vt.it

**Summary**

Fantappié-Arcidiacono projective relativity, also known as "de Sitter relativity", provides a formulation of mechanics according to which it is possible to identify an inertial frame of reference by means of kinematic methods. This article presents the hypothesis that inertial frames of reference originated during the so-called "archaic era" of the Universe. A simple mechanism is proposed for this process. Such a process could leave behind cosmic fluid granulation, which could manifest itself in the present Universe as a clustering of dark matter in galaxies.

## 1. Introduction

In a series of previous works [1,2,3], the concept of a phase in the history of the Universe which "preceded" the big bang was introduced in the reference of Fantappié-Arcidiacono projective relativity (also known as de Sitter relativity) [4,5,6]: the so-called "archaic era". While referring to the original works for a more detailed exposition, here we shall confine ourselves to recalling some key concepts.
The archaic Universe was a sort of quantum vacuum populated solely by virtual processes - real particles did not exist. The geometry of this Universe was that of the surface of a 5-dimensional Euclidean sphere, the so-called Arcidiacono hypersphere. A privileged direction, corresponding to a temperature axis, existed on this surface.
A special maximum circle perpendicular to this direction corresponded to an infinite temperature value. From now on we shall refer to it as "the equator". The equator divided the surface of the hypersphere into two separate semispheres. The circles parallel to the equator located on one of the two semispheres corresponded to curved three-dimensional spaces having temperature values which decreased as the distance from the equator increased.
The semisphere was the locus of quantum fluctuations whose initial extreme was located on the equator and whose final extreme was located on the circle corresponding to a definite temperature. This background of fluctuations satisfied the uncertainty principle and was homogeneous and isotropic in the sense that its local properties depended solely on the temperature and not on the spatial position or direction.

The archaic era ended when the temperature decreased to a value equal to the hadrons stabilization energy. At that point, a sort of phase transition occurred which made the surviving fluctuations real; this transition was the big bang. Starting from that moment, the dynamics of the elementary components of matter was no longer described in terms of the coordinates on the "archaic" chronotope, but in terms of coordinates derived from these by Wick rotation, projection and scale reduction. Thus the initial hyperdense plasma appeared, whose evolution was subject to the ordinary equations of state and to the gravitational equations of Arcidiacono projective general relativity (PGR).

PGR differs from Einstein's ordinary general relativity in that its limit case for null matter-energy density (in the absence of the cosmological term) is not the Minkowski chronotope but the projective version of the de Sitter chronotope (so-called Castelnuovo chronotope). In this limit case, PGR becomes the projective special relativity (PSR), developed by Fantappié and Arcidiacono as far back as 1954.

The essential difference between PSR and the usual Einstein special relativity is that, in addition to the limit propagation velocity $c$ (already present in Einstein's theory), the limit duration $t_0$ (that is the chronological distance of *each* observer from the de Sitter horizon) also appears in it [4,7,8]. The de Sitter radius $r = ct_0$ coincides with the radius of the archaic hypersphere in 5-dimensional space. Two new fundamental constants of Nature thus appear in PSR, and not just one.

From a physico-mathematical point of view, PSR mechanics appears interesting for at least two reasons: 1) the generalized inertia tensor combines both the usual moment of inertia and the static moment and the mass, in a surprising unification [4,9]; 2) it is possible, by means of internal procedures of a kinematic type, to test if a given frame of reference is inertial or not. These results lead to believe that the problem of the nature and origin of inertia can be posed in a new light if it is addressed using concepts and interpretational tools taken from PSR [9].

Indeed, it is important to underline that the PSR inertial frames of reference are global, and that the transformation from one of these frames to another belongs to a global group of transformations (the de Sitter group), while the usual Einsteinian strategy allows a curved spacetime only by giving up global inertial frames of reference [4,5,7,8]. The equivalence principle, however, remains valid owing to its *local* nature [2,3].

In Section 2 brief reference is made to the PSR results concerning the "fixed stars frame of reference"; for historical reasons we shall use this outdated term instead of the more current "Blackbody Background Radiation frame of reference", the cosmocentric frame of reference where the BBR dipole anisotropy is null. In Section 3, the connection between this frame and the quantum vacuum of the "archaic era" is addressed. The latter was homogeneous and isotropic in a very special frame of reference. At the big bang, this frame became a global inertial frame of reference. Through this hypothesis, the origin of inertia is derived from the uncertainty principle.

Though PSR is the most general relativistic theory, based on a group of global transformations, compatible with the inertia principle [1,10,11,12], it is possible that the latter admits of exceptions. In Sections 4-6 the possibility is discussed that the "cosmic fluid", utilized to describe the continuum of the fundamental frames of reference in cosmology, has a granular structure inherited from the quantum fluctuations of the archaic era. Within each individual granule or "molecule" of fluid there would seem to be a slight breach of the inertia principle and this would give rise to effects (currently not modelled) which could explain the observed anomalies in the profiles of galactic rotation velocities without resorting to hypotheses such as the presence of clustered dark matter.

Section 7 touches on a possible, and still conjectural, type of solution to the problem of the dominance of matter over anti-matter.

## 2. PSR and "fixed stars" frame of reference

The so-called "fixed stars frame of reference" is defined through appropriate average operations on the motion of celestial bodies; its inertiality represents an empirical fact of a cosmological nature, but it is not explained by classical mechanics. The latter confines itself to postulating the existence of at least one inertial frame of reference with respect to which Newton's laws of motion can be formulated; starting from this, an infinity of other inertial frames of reference can be obtained by means of Galileo transformations.

In PSR, cosmology is strictly connected with mechanics, so that it is possible to formulate an entirely mechanical (indeed, kinematic) procedure for the identification of an inertial frame of reference. A frame of reference, connected with a certain observer, is inertial when all the celestial bodies recede from that observer with a velocity (averaged out on the peculiarities of the motion), which is independent of direction and increasing with the distance from him according to a very precise law (de Sitter expansion). In the limit $t_0 \to \infty$, this velocity becomes identically zero, and the empirical result valid for classical mechanics is recovered (see ref. [9], in particular eqs. (4.16) and following).

## 3. Archaic quantum mechanics and thermal physics

The equation of the Arcidiacono 5-sphere in projective coordinates is the following:

$$\bar{x}_\mu \bar{x}^\mu + \bar{x}_5 \bar{x}^5 = r^2 \qquad . \qquad \mu = 0, 1, 2, 3. \qquad (1)$$

The equation of PGR space connected with a single fundamental observer is the following [2]:

$$\bar{x}_0 \bar{x}^0 - R^2(\tau) \bar{x}_i \bar{x}^i + \bar{x}_5 \bar{x}^5 = r^2 \qquad . \qquad i = 1, 2, 3. \qquad (2)$$

Equation (2) can be imagined as being derived from equation (1) through a Wick rotation making the time intervals real, followed by a contraction of spatial coordinates by a scale factor $R(\tau)$, dependent on cosmic time $\tau$. The inverse of this rotation can be expressed as ($k$ = Boltzmann's constant, $\hbar$ = reduced Planck's constant, $T$ is real and positive or null):

$$t \to \frac{j\hbar}{kT}, \qquad j = \sqrt{-1}, \qquad (3)$$

from which it follows that:

$$\frac{dT}{dt} = \frac{jkT^2}{\hbar} .$$

Let us now consider equation (2) in an infinitesimal neighbourhood of the observer ($\bar{x}_\mu = 0$, $\bar{x}_5 = r = ct_0$). Let us consider, in this neighbourhood, a free non-relativistic particle having a mass $m$ described by a wavefunction $\Psi$ which satisfies the usual Schrödinger equation:

$$\frac{\hbar^2}{2m}\Delta\Psi = j\hbar\frac{\partial\Psi}{\partial t} \quad . \tag{4}$$

The spacetime coordinates which appear in equation (4) are, of course, the physical coordinates associated with the projective coordinates normalized in accordance with equation (2). Let us now define the absolute "temperature" $T$ through the substitution (3). We obtain:

$$\frac{\hbar^2}{2m}\Delta\Psi = j\hbar\frac{\partial\Psi}{\partial T}\frac{dT}{dt} = -kT^2\frac{\partial\Psi}{\partial T} \quad . \tag{5}$$

Equation (5) can be considered the "ancestor" of equation (4) which is valid on the archaic space whose coordinates satisfy the normalization condition (1). For the sake of simplicity, we shall not consider here the inverse operation of the scale contraction by a factor $R(\tau)$.

Assuming, therefore, that equation (5) represents the equation of motion of a free particle in a very small neighbourhood of archaic space, and confining ourselves for simplicity's sake only to the one-dimensional case, let us seek separable solutions of the form:

$$\Psi(x,T) = \varphi(x)\Lambda(T) \quad . \tag{6}$$

By easy passages one has:

$$\Lambda = const \times \exp(\lambda/kT)$$

$$\frac{d^2\varphi}{dx^2} = \frac{2m\lambda}{\hbar^2}\varphi \quad . \tag{7}$$

Posing $\varphi = \exp(jfx)$ one has:

$$\frac{2m\lambda}{\hbar^2} = -f^2 \quad .$$

Let us consider before the possibility $\lambda < 0$. In this case, we can pose $\lambda = -E$ with $E > 0$ and this results in:

$$f = \pm\frac{\sqrt{2mE}}{\hbar}$$

being real. The required solution thus becomes:

$$\Psi = \Psi_0 \exp\left[\pm j\frac{\sqrt{2mE}}{\hbar}x - \frac{E}{kT}\right] \quad . \tag{8}$$

Though a rigorous treatment of the behaviour of this solution outside of a neighbourhood would require the study of de Sitter-covariant equations, one can easily infer on a physical basis that, since archaic space is closed, the spatial coordinates are cyclical with a period $2\pi r$. Thus, from the monodromy of equation (8) the following quantization rule is derived:

$$\frac{\sqrt{2mE}}{\hbar} \times 2\pi r = 2\pi n \quad \Rightarrow \quad E = \frac{n^2 \hbar^2}{2mr^2}; \quad n=0,1,2,... \qquad (9)$$

That is, the kinetic energy is quantized as a function of the polar moment $mr^2$ of the particle.

Let us now consider the alternative possibility $\lambda > 0$; in this case $f = j\sigma$ where $\sigma$ is real and $\varphi = \exp(jfx) = \exp(-\sigma x)$. If $\sigma > 0$, $\varphi \to 0$ for $x \to \infty$ and $\varphi \to \infty$ for $x \to -\infty$; if $\sigma < 0$ the opposite applies. This solution cannot be periodic and therefore cannot be defined as an one-valued function on closed space; it must therefore be ruled out as not physical.

Therefore, the only possible solution with separable variables is that given by equation (8), and it is easily seen that it is not a *T*-stationary solution (as we are dealing with a "virtual" fluctuation). By applying the inverse of substitution (3), equation (8) becomes:

$$\Psi = \Psi_0 \exp\left[\pm j \frac{\sqrt{2mE}}{\hbar} x - j \frac{Et}{\hbar}\right] \qquad (10)$$

i.e. a progressive plane wave having energy *E*. We think that this is the transformation which $\Psi$ undergoes in the transition constituted by the "big bang".

What is the meaning of equation (8) during the archaic era, i.e. "before" its transformation into equation (10) comes about? [Naturally, the concept of "before" is to be understood here as conventional, we rather ought to say: "for temperatures *T* exceeding the transition temperature"].

Let us assume the validity of the Born rule during the archaic era. Thus the total probability of the particle being located in an arbitrary point of archaic space will be:

$$P = \int dV \|\Psi\|^2 = \exp\left(-\frac{2E}{kT}\right), \qquad (11)$$

where $V = 2\pi^2 r^3$ is the total (finite) volume of archaic space and $(\Psi_0)^2 = V^{-1}$. This relation can easily be rewritten as:

$$\Delta S = \frac{\Delta Q}{T} \qquad (12)$$

posing:

$$\Delta S = k \ln P; \quad \Delta Q = -2E. \qquad (13)$$

At temperature $T = \infty$, i.e. on the equator of the Arcidiacono hypersphere, one has $P = P_0 = 1$ and therefore $S_0 = k\ln P_0 = 0$. Thus defining, for an arbitrary $T$, $\Delta S = S(T) - S_0 = S$ the first equation (13) becomes the usual Boltzmann relation $S = k\ln P$.

As can easily be seen from equation (11), the probability of the existence of the particle having energy *E* is not conserved as the temperature varies. A net exchange of kinetic energy thus occurs between the particle and the heat bath at the temperature *T*; this exchange also involves the rest energy of the particle, which, however, is not considered here because we have adopted a non-relativistic wavefunction. The second equation (13) describes this heat exchange[1]. It allows the

---
[1] The factor 2 in the Helmholtz free energy expression 2*E* derives from the fact that in the archaic era no transaction [13] has yet been initiated, and therefore the heat bath supports both the "delayed" particle with energy *E* and the

Born rule to be expressed, through equation (12), in the form of the Clausius equality for reversible thermodynamic processes. This observation allows to give a thermodynamic meaning to the archaic wavefunction (8): archaic quantum mechanics is a form of thermostatics.
By dividing the two terms of equation (12) by $k$, one has:

$$\frac{\Delta S}{k} = \frac{\Delta Q}{kT} = \frac{-2E}{kT} \quad ;$$

and by applying the inverse of substitution (3) this expression becomes $2Et/\hbar$. One is therefore dealing with the action-entropy correspondence relating to fluctuations originating at $T = \infty$ and ending at temperature $T$. At this temperature they are reabsorbed in the heat bath, giving rise to equation (12).

The evanescent waves represented by equation (8) have finite amplitude on the equator of the Arcidiacono sphere ($T = \infty$); this amplitude vanishes in the limit $T \to 0$. The "free particles" of the archaic era are basically vacuum virtual fluctuations originating on the equator and ending at the parallel corresponding to temperature $T$. These waves are static (there are no imaginary phase factors dependent on $T$ in equation (8)), which means that they are "at rest" in the frame of reference with respect to which equation (8) is formulated. When the big bang happens these waves turn into real particles described by equation (10). These new waves satisfy the usual relativistic dispersion relation between wavelength and frequency, in which the invariant "particle rest mass" appears. This is the origin of mass.
The spacetime frames of reference obtained by applying the inverse of substitution (3) to the archaic frame of reference within which equation (8) is valid are inertial frames. It is in this way that inertial frames of reference originate.
Thus, inertia emerges as the effect of the existence, in the archaic era, of a frame of reference with respect to which virtual fluctuations are described by equation (8) without imaginary phase terms dependent on temperature. The existence of such a frame is obvious because, otherwise, the generic fluctuation would have a defined energy and motion and therefore would not be virtual[2]. In other words, this frame exists because the archaic era is populated only by virtual fluctuations, to which the uncertainty principle applies without exception.

## 4. The "cosmic fluid" and its molecules

The concept of "cosmic fluid" is often used in usual relativistic cosmology; it has been also applied to the formal development of PGR based on equation (2) and on the cosmological principle [2,3]. This fluid represents the substratum of fundamental inertial observers and is therefore continuous: its "molecules" are assumed to be lacking in spatial extension. Let us now investigate a possible complication, consisting in the assumption that the cosmic fluid, though continuous, presents as inclusions "molecules" of finite extension. While the centroid of the individual molecule must necessarily still coincide with the origin of an inertial frame of reference, this might no longer hold true for the other internal points of the "molecule".
We shall firstly place ourselves in the simplest case in which $R \equiv 1$, i.e. in the PSR context. A "molecule" shall be a cosmic fluid flow tube, constituted by a continuous sheaf of fundamental

---

"advanced" particle having identical kinetic energy. When a transaction begins, these two "particles" become the delayed and advanced part of the propagation of a single particle having energy $E$. This topic is better expressed in a work to be published soon.

[2] More precisely, the phase term dependent on $T$ could be decomposed into a sum of harmonic terms corresponding to an equal number of fluctuations having a definite energy.

observers. However, we shall renounce to the idea that these observers, though at rest with respect to the inertial frame of reference having as its origin the centroid of the tube, are inertial.

Thus, a free material point initially at rest with respect to one of these observers and placed in its same position will be accelerated with respect to the centroid. We shall assume that the successive free motion of this material point is periodic, with a period $p$ subjected to the quantization rule:

$$(\hbar c/r)p = v\, e^2/c \qquad\qquad v = 0, 1, 2, ... \qquad\qquad (14)$$

where $e$ is the elementary charge. I.e.:

$$cp = v\alpha r; \qquad v = 0, 1, 2, ...; \qquad \alpha = \frac{e^2}{\hbar c} \quad .$$

Equation (14) can be partially justified by assuming the molecules originate during the archaic era; this issue will be discussed in greater depth in the next Section. The "duration" of the archaic era is equal to $e^2/mc^3$, where $m$ denotes the mass of the electron [1,2,3]. Since the electron is the lightest particle among those having a defined mass (neutrinos are lighter, but are created by weak interactions in oscillating states which are superpositions of mass eigenstates), it is possible to associate with the archaic era a quantum of elementary action equal to the product of the electron's rest energy $mc^2$ and the "duration" $e^2/mc^3$; this quantum is $e^2/c$. On the other hand, the left member of equation (14) can be written in the form $\hbar p/t_0$; it is equal to $\hbar$ when the oscillation period $p$ is equal to the distance $t_0$ of the de Sitter horizon. The precursor of the oscillation during the archaic era is a virtual process and the action associated with this process must be less than or equal to $\hbar$; it is precisely equal to $\hbar p/t_0$. Equation (14) is therefore the expression of the assumption that this action is a multiple of the quantum of action $e^2/c$. If one takes into account the fact that $\Delta v = \pm 1$, equation (14) can be written in the alternative form $\Delta(\hbar p/t_0) = e^2/c$ and in this form is particularly suggestive, as the increase in action coincides with the value $e^2/c$ which in the classical theory of electrically charged elementary particles is associated with the breakdown of the inertia principle[3].

Since the period $p$ is measured in terms of the proper time of the centroid, equation (14) can be considered a quantization rule of the action of an oscillator in the centroid rest frame of reference. Consequently, $\hbar c/r$ is the rest energy of the oscillator associated with our test material point placed inside the molecule (and must not be confused with the rest energy of the point, nor with that of the molecule!). In other words, the mass of this oscillator is $\hbar/rc$.

Now, equation (14) tells us that $p$ does not depend on the spatial position of the point, nor on the amplitude of the periodic motion. This means that the motion is harmonic and therefore the oscillation amplitude $x$ satisfies the differential equation:

$$\ddot{x} + \omega^2 x = 0 \quad .$$

The statement that the material point is free means that the force exercised on the oscillator associated with it does not exceed, in modulus, the "minimum force". This, on the PSR chronotope, is the ratio of the least allowed variation of the impulse (i.e. $\hbar/r$) to the maximum time interval that can be assigned to such a variation (i.e. $t_0$). The maximum acceleration that can be experienced by the test point with respect to the centroid is therefore:

---

[3] If it is required that the radiation reaction be very small, i.e. that $(e^2/c^3)\Delta a/\Delta t \ll ma$, where $m$ is the mass of the charged particle and $a$ its acceleration, also when $\Delta a/a \gg 1$, one immediately obtains that the action increase $mc^2\Delta t$ must be $\gg e^2/c$.

$$a_{max} = \frac{rc}{\hbar} \frac{\hbar/r}{t_0} = \frac{c}{t_0} ,$$

i.e. that which in a previous work we have called "reference acceleration" [3].
This acceleration will be associated to the maximum amplitude $\rho$ such that:

$$\omega^2 \rho = a_{max} = \frac{c}{t_0} \quad \Rightarrow \quad \omega = \sqrt{\frac{c}{\rho t_0}} .$$

I.e.:

$$\omega = \frac{2\pi}{p} = \sqrt{\frac{c}{\rho t_0}} . \tag{15}$$

This relation can be written as:

$$\rho = \frac{(cp)^2}{4\pi^2 r} .$$

It thus follows from equation (14):

$$\rho = \frac{v^2 \alpha^2 r}{4\pi^2} . \tag{16}$$

Since the problem is spherically symmetrical (the maximum amplitude $\rho$ does not have a directional dependence, as can be inferred from equation (16)), the molecule is spherical with a radius $\rho$. Equation (16) thus becomes a quantization rule for the radius.
Since the centroid is inertial, the rest position of the oscillator ($x = 0$) must coincide with it. In conclusion, a material point placed at the distance $x$ from the centre O of the molecule undergoes a radial acceleration towards O equal to:

$$\mathbf{a} = -\frac{c}{t_0} \frac{\mathbf{x}}{\rho}; \quad 0 \le x \le \rho$$

$$\mathbf{a} = 0; \quad x \ge \rho . \tag{17}$$

This result clarifies the meaning of the period $p$. If the test material point is released with null initial velocity, it moves along a trajectory which passes through the centre and passes through it again at intervals of duration $p$. Therefore, the test point will coincide with the origin of an inertial frame of reference only at instants separated by intervals of duration $p$.
The potential energy per unit mass is therefore:

$$U = -\frac{c}{2\rho t_0} x^2 . \tag{18}$$

For a typical hadron this energy exceeds $10^8$ °K, which is well lower than big bang temperature according to the present model ($\approx 10^{13}$ °K); therefore during the archaic era the influence of inertia fluctuations on the matter was very small.

Since $\rho_{max} = r$ one has $v_{max} = 2\pi/\alpha \approx 6 \times 137$. The smallest non null value of $\rho$ is instead

$$\rho_{min} = \frac{\alpha^2 r}{4\pi^2} \approx 10^{-6} r .$$

The case $v = 0$, $\rho = 0$ corresponds to the customary molecule having null spatial extension. In the space uniformly filled by this type of molecules alone, the inertia principle is strictly valid everywhere.

Let us suppose a molecule A having radius $\rho_1$ to be included in a molecule B having radius $\rho_2 > \rho_1$. A test point placed within the molecule A undergoes to an acceleration $a_B$ with respect to the centre of B which must be related, in some way, to the acceleration $a_A$ of the same point with respect to the centre of A. In line of principle, two composition rules can be considered:

1) At the position where the test point is placed two several acceleration fields $a(\rho_1) = a_A$ and $a(\rho_2)$ exist, both of which are expressed by eqs (17); $a_B$ is then given by the vectorial sum $a(\rho_1) + a(\rho_2)$.

2) Only the centre of B is really inertial, while the centre of A is accelerated towards the centre of B according to eqs (17) [the initial speed of A with respect to the centre of B is not necessarily null]. $a_B$ is the sum of $a(\rho_1) = a_A$ and this second acceleration. As an effect of the space expansion, A and B sizes growth until B size is equal to $\rho_2$; after that, A and B sizes are no more influenced by the expansion.

The first possibility is contemplated in rif. [3], eq. (16). However, the second possibility is probably more realistic. It is equivalent to assume the free motion of the test point is multiperiodic, with frequencies related to radii $\rho_1$, $\rho_2$. The argument is easily generalized to a chain of molecules A, B, C, ... M of different radii.

## 5. The pre-cosmic origin of the "molecules"

If the generalization suggested in the previous Section is valid, a PSR fundamental observer could be not genuinely inertial. A free body would experience, in the origin of a reference solidary with this observer, an acceleration $\mathbf{a}$ of modulus $\leq c/t_0$; the deviation from inertiality would be expressed by the field $\mathbf{a}$. What might have been the precursor of this acceleration in the archaic era?

Let be $x$ the distance of a material point from the centre of a molecule; when the transform (3) is applied, its acceleration $d^2x/dt^2$ towards the centre of the molecule becomes:

$$\frac{d^2x}{dt^2} \rightarrow \frac{d^2x}{\left(\frac{dt}{dT}dT\right)^2} = -\frac{k^2 T^4}{\hbar^2}\frac{d^2x}{dT^2} . \tag{19}$$

In the archaic era the Wick-rotated PSR was valid and locality did not hold, because time did not exist [1,2]. We can suppose that the molecules appeared and dissolved as special non-local virtual fluctuations subject to the uncertainty principle. Let us consider the fraction of molecules arrived at the instant of the big bang; these molecules became acceleration fields articulated in (superposed) spherical molecules of the type described by equations (17). These molecules, whose number was finite because their total energy had to be finite, thus became "condensation nuclei" immersed in a three-dimensional continuum of oscillators $v = 0$, $\rho = 0$ having null energy (inertial substratum). We note that the scale factor $R(\tau)$ does not act on $\rho$, so that the single acceleration field represented by equation (17) is not affected by the expansion of space. This latter has only the effect to separate the centres of different molecules.

As it has become real and is no longer made up of virtual fluctuations, this cosmical field of accelerations left over after the big bang remained frozen: it survives to the present day. The distance between the centres of the various molecules increased in time solely as an effect of the scale factor $R(\tau)$. With the big bang the real interactions, for example gravitational ones, began. At this point the condensation phenomena considered in conventional models started, conditioned however by the presence of the condensation nuclei (17). Thus a fluid of nested condensation "cells", homogeneous and isotropic on a large scale, was formed.

The typical size of the cell is $2\rho$. If we indicate with $H_0$ the value of the Hubble parameter in the current era we obtain a "quantum of expansion velocity", associated with the cell size $2\rho$, equal to $2\rho H_0$. The optimal values of $t_0$ and of $H_0$ have been estimated in a previous work [3], and turn out to be, respectively, $4.822 \times 10^{17}$ s and $1.67 \times 10^{-18}$ s$^{-1}$. If one considers the values of $v$ around 5 (see Table 1) and takes into account the uncertainty about $H_0$, this "quantum of velocity" is actually in the same range of "velocity quantization" observed by Tifft in galactic clusters [14]. This "quantization" could therefore be an effect of statistical prevalence generated by the presence of the condensation nuclei (17). Indeed, by observing a sufficiently compact cluster of cells, the distribution of the distance intervals of the centers of adjacent cells, along any line of sight whatsoever, will show a peak at the value $2\rho$. The distribution of redshift intervals measured on these cells will therefore have a maximum at the value $2\rho H_0$. It is obvious, however, that this "quantization" is easily destroyed by the peculiar motions within the cluster and by the insufficient compactness.

| Table 1 | |
|---|---|
| $v$ | $2\rho H_0$ (km s$^{-1}$) |
| 3 | 5.82 |
| 4 | 10.35 |
| 5 | 16.17 |
| 6 | 23.29 |
| 7 | 31.70 |

### 6. Galactic dark matter

It is reasonable to believe that galaxies correspond roughly to "molecules" of cosmic fluid, in that they derive from the condensation of primordial matter around molecular centers. The values of the $p$ and $\rho$ parameters estimated for $v = 2$ ($7 \times 10^{15}$ s and $8 \times 10^{20}$ m, respectively, if the value of $t_0$ indicated previously is adopted) are very similar to the rotation period and radius of the Milky Way. A free material point placed inside a molecule having a radius $\rho$ at a distance $x$ from the centre will be in mechanical equilibrium under the action of centrifugal force if:

$$\frac{v^2}{x} = \frac{c}{t_0} \frac{x}{\rho} \quad , \tag{20}$$

where $v$ is the revolution velocity of the point. Thus, in the absence of self-gravitation, the equilibrium velocity is proportional to the distance from the centre, up to a maximum value at the distance $x = \rho$.

If, rather than considering a free material point, one considers an extended distribution of mass having spherical symmetry with a mass function $M(x)$ [mass contained within the distance $x$ from the centre], the equilibrium condition becomes:

$$\frac{GM(x)}{x^2} + \frac{c}{t_0}\frac{x}{\rho} = \frac{v^2}{x} \quad . \tag{21}$$

The same equilibrium acceleration profile can be obtained by self-gravitation alone if a mass function $M_0(x)$ is hypothesized such that:

$$\frac{GM_0(x)}{x^2} = \frac{v^2}{x} \quad . \tag{22}$$

It is easily deduced from equations (21), (22) that:

$$\frac{M_0(\rho) - M(\rho)}{M_0(\rho)} = \frac{c/t_0}{v^2(\rho)/\rho} \quad . \tag{23}$$

If the distance $x = \rho$ is roughly identified with that at which the radial velocity profile of the galaxy is at its maximum, for many spirals the right-hand member of equation (23) is near to 1. Such is the case, for example, of NGC 801, NGC 3198 and NGC 6503. The observed distance $\rho$ is in the same range of $\rho_{min}$ (6.3 kpc).

This implies, as can easily be seen from equation (23), that $M(\rho) \ll M_0(\rho)$. In other terms, if in the condition of mechanical equilibrium one omits the term associated with the violation of the inertia principle due to the immersion of the galaxy in the molecule of cosmic fluid, a highly overestimated mass function is derived from the interpretation of the galaxy velocity profile. This is a possible explanation for clustered dark matter in galaxy systems. Methods of evaluation of galactic dark matter different from those based on the rotation velocity profile of galaxies (e.g. gravitational micro-lensing) will not be discussed here.

Let us recall that as far as the dark matter spread throughout intergalactic space is concerned, on the other hand, projective relativity allows an explanation of it as an artefact of global curvature [3].

## 7. Two mutually specular worlds?

In all the reasoning set forth in this article it has been tacitly assumed that matter distributed in space is ordinary matter or, in other terms, that the percentage of antimatter is negligible.

The reason for the dominance of ordinary matter in our Universe remains an enigma also within our approach. The equator of the Arcidiacono hypersphere corresponding to an infinite absolute temperature is the origin of all the "archaic" quantum fluctuations. Those which end at a temperature equal to the creation energy of nucleons are converted into real particles which emerge as the "big bang". Since from this moment on the temperature is too low for there to be interconversion of baryons and antibaryons, the dominance of matter must be established beforehand, during the "archaic era".

However, in the archaic era only the quantum vacuum is present, which as such is expected to be symmetrical to the utmost. The enigma, therefore, reappears in this approach, as well.

It must be borne in mind, however, that in the approach presented here it is possible to exploit a symmetry which is lacking in more conventional cosmology. Our Universe develops from one of the two semispheres into which the infinite-temperature equator divides the Arcidiacono hypersphere. The semisphere is converted into the PGR chronotope by means of a Wick rotation and the application of a scale reduction.

The other semisphere does not play any role and can be considered as a simple mathematical artefact (it corresponds to the choice of a negative sign instead of a positive sign in the definition of the projective coefficient $A$ of the metrics [2,4,9]). However, it is also possible that this semisphere corresponds to a second archaic universe that is specular with our own. If the archaic fluctuations exiting from the equator towards this second semisphere corresponded solely to antimatter particles and those exiting towards our semisphere corresponded solely to particles of ordinary matter, the symmetry of the original vacuum would be complied with.

Thus, in the second semisphere, by means of the same mechanisms seen above, a Universe would be developed in which antimatter would be dominant. And this would take place whilst maintaining the symmetry of the archaic quantum vacuum. These two "mutually specular Universes" would be separated by the equator, and would therefore be causally unconnected, albeit contiguous. Their common origin would be the archaic quantum vacuum.

The mechanism based on which ordinary matter and antimatter would seem to separate along the equator remains enigmatic. A particle differs from its antiparticle by the sign of the charges with which they couple with the various fields. On the other hand, if one admits that the various interaction fields present in the archaic era satisfy the Gauss theorem, one easily sees that the charges must be null because of the fact that space is closed [15]. One can therefore presume that the charges make their appearance only at the time of the big bang, when the particles become real and therefore capable of real interaction and space becomes open. Only at that time does each particle acquire its own charge.

If one assumes that the sign of the charges is defined by the semisphere in which the particle becomes real (i.e. by the fact that it appears at the time of the big bang in our Universe or, alternatively, at the time of the anti-big bang in the anti-Universe) one does indeed obtain the required separation *ab initio* of matter and antimatter. The sign of the charge would in other words be defined by the direction of the timeline emerging from the equator along which the particle materialized, a result which could in some way be connected with the CPT theorem.

In addition, it is possible to assume that within the antimatter semisphere eq. (3) holds with a sign minus at left hand. By substituting it in the advanced form of Schrödinger equation, that is the complex conjugate of eq. (4), eq. (5) is derived again. However, eq. (10) is now converted in the corresponding advanced wavefunction. In other terms, the fluctuation (8) of energy $E > 0$ which before was converted in the retarded wavefunction (10) of energy $E > 0$, now is instead converted in the advanced wavefunction represented by the complex conjugate of (10), having an energy $-E < 0$. In this way, a negative energy can be associated with quantum fluctuations on the antimatter semisphere. Due to the simmetry of initial state at infinite temperature, this negative energy exactly compensates the positive energy of fluctuations on the matter semisphere, so enabling a *creatio ex nihilo*. However, for an inertial observer exiting from the anti-big bang in the anti-Universe, which measures a growing cosmic time ($t \rightarrow -t$), the energy released at big bang in form of antimatter is positive.

Though the entire conjecture does not appear open to verification by observation, it nevertheless allows to explain the dominance of ordinary matter without introducing special initial conditions, at the same time preserving the greatest simmetry of the initial state.

In line of principle, another possibility is that separation *ab initio* has never happened, but instead a quantity A (B) of matter (antimatter) has been injected in our Universe at the big bang, and a quantity A (B) of antimatter (matter) has been injected in the anti-Universe at the anti-big bang. In this case, the condition A > B assures the dominance of matter in our Universe, which is however balanced by the dominance of antimatter in the anti-Universe. The total annihilation of antimatter in early days of our Universe should have released a large amount of radiation pressure [16].

**8. Conclusions**

This article aims to propose, within the context of Fantappié-Arcidiacono projective relativity, a theoretical justification of that which appears to be an ascertained empirical fact, i.e. that the fixed stars frame of reference is inertial (just think of the rotation of the oscillation plane of a Foucault pendulum at the Pole...).
We ask ourselves: from what cosmological situation does a global inertial frame of reference evolve in which the stars appear fixed? As we have seen in Section 2, PSR seems to be an excellent starting point to solve this type of problem. It would already be an answer in itself, if it were not that in it the Universe lacks matter (this fact, however, confirms that inertia is not associated with the relative motion of bodies, contrarily to Mach's argument [17]). A true answer, therefore, must be sought in PGR-based cosmology.
To a PGR fundamental observer the initial singularity of the big bang must appear ever more flattened on the de Sitter horizon with the passage of cosmic time. Indeed, only in this way can the intersection between the time axis of the observer with such a horizon and a time origin of that observer in the big bang exist simultaneously. This implies that the big bang cannot be pointlike in an absolute sense, but only from the individual observer's point of view. The existence of an absolute geometric reference (a modern edition of Milne's "public space" [18]) in which the big bang is extended leads directly to a pre-big bang phase, which is the archaic era.
As Licata acknowledged as far back as 1994 [5,19,20], the archaic era Universe is basically a form of quantum vacuum populated only by virtual fluctuations. The phase transition which closes the archaic era, i.e. the big bang, creates an inertial cosmic frame of reference as a legacy of the archaic frame of reference with respect to which this vacuum was empty, i.e. was homogeneous and isotropic. It is with respect to this frame of reference that physical laws such as Schrödinger's equation or Newton's second law are formulated.
In the limit in which the density of matter tends to zero, PGR is converted into PSR [2], and therefore the PGR inertial frame of reference collapses on the PSR inertial frame of reference, in which the average motion of the "stars" is null and only de Sitter kinematic expansion survives [9]. A further limit for $t_0 \to \infty$ makes this expansion disappear and leads to the "fixed stars frame of reference" used in classical physics. Thus the cosmic inertial frame of reference coincides with the "fixed stars frame of reference". The fact that the fixed stars frame of reference is inertial does not derive, therefore, either from the existence of a substance pervading all of space and which would be immobile in such a reference (aether) or from the absence of relative motion with respect to the stars (as argued by Mach with reference to the rotating bucket experiment [17]). Rather, "absolute space" would be a relic of the archaic era "preceding" the big bang.
If this theory is correct, it is possible that the archaic quantum fluctuations have released in fossil form volumes of space where the inertia principle is slightly violated. Matter would tend to condense inside these regions, and it seems plausible that galaxies and clusters represent hierarchies in condensation phenomena conditioned by this effect.
Some effects attributed to dark matter clustered in galaxies or within clusters might be caused by such subtle violations.

**Acknowledgments**

The author wishes here to thank Ignazio Licata for many stimulating discussions and Elmo Benedetto for the constant encouragement.